\documentclass{aa}
\usepackage{graphicx}
\usepackage{txfonts}
\begin{document}
   \title{The envelope mass of red giant donors in type Ia supernova progenitors}


   \author{Xiangcun Meng
          \inst{1}
          and
          Wuming Yang \inst{1}}

   \offprints{X. Meng}

   \institute{School of Physics and Chemistry, Henan Polytechnic
University, Jiaozuo, 454000, China\\
              \email{xiangcunmeng@hotmail.com}
             }

   \date{Received; accepted}


  \abstract
   {The single degenerate model is the most widely accepted progenitor model
   of type Ia supernovae (SNe Ia), in which
   a carbon-oxygen white dwarf (CO WD) accretes hydrogen-rich
   material from a main sequence or a slightly evolved star (WD +MS)
   or from a red giant star (WD + RG), to increase its mass and explodes when
   approaching the Chandrasekhar mass. The explosion
   ejecta may impact the envelope of and
   strip off some hydrogen-rich material from the companion. The
   stripped-off hydrogen-rich material may manifest itself by
   means of a hydrogen line in the nebular spectra of SNe Ia.
   However, no hydrogen line is detected in the nebular spectra.}
   {We compute the remaining amounts of hydrogen
 in red giant donors to see whether the conflict between theory and observations
 can be overcome.
}
   {By considering the mass-stripping effect from
   an optically thick wind and the effect of thermally unstable disk,
   we systematically carried out binary evolution calculation for WD + MS and WD + RG systems.
}
   {Here, we focus on the evolution of WD + RG systems. We found that some donor stars at the time of the supernova explosion
   contain little hydrogen-rich material on top of the helium core (as low as 0.017 $M_{\odot}$),
   which is smaller than the upper limit to the amount derived from observations of material stripped-off by explosion ejecta.
   Thus, no hydrogen line is expected in the nebular spectra of these SN Ia.
  We also derive the distributions of the envelope mass and the core mass of the companions from WD + RG channel at the
  moment of a supernova explosion by adopting a binary population synthesis approach. We rarely find a RG companion with a very low-mass envelope.
  Furthermore, our models imply that the remnant of the WD + RG
  channel emerging after the supernova explosion is a single low-mass
  white dwarf (0.15 $M_{\odot}$ - 0.30 $M_{\odot}$).}
   {The absence of a hydrogen line in nebular spectra of SNe
   Ia provides support to the proposal that the WD + RG system is the progenitor of SNe Ia. }

   \authorrunning{Meng \& Yang}
   \titlerunning{The envelope mass of red giant donors in type Ia supernova progenitors}
   \maketitle{}
%

\section{Introduction}\label{sect:1}

Although type Ia supernovae (SNe Ia) are clearly important in
determining cosmological parameters, e.g., $\Omega_{\rm M}$ and
$\Omega_{\Lambda}$ (Riess et al. \cite{REI98}; Perlmutter et al.
\cite{PER99}), the progenitor systems of SNe Ia have not yet been
confidently identified (Hillebrandt \& Niemeyer \cite{HN00};
Leibundgut \cite{LEI00}). It is widely believed that a SN Ia is
produced by the thermonuclear runaway of a carbon-oxygen white
dwarf (CO WD) in a binary system. The CO WD accretes material from
its companion to increase its mass. When its mass reaches its
maximum stable mass, it explodes as a thermonuclear runaway and
almost half of the WD mass is converted into radioactive nickel-56
(Branch \cite{BRA04}). Two basic scenarios have been presented.
One is the single degenerate (SD) model, which is widely accepted
(Whelan \& Iben \cite{WI73}; Nomoto et al. \cite{NTY84}). In this
model, a CO WD increases its mass by accreting hydrogen- or
helium-rich matter from its companion, and explodes when its mass
approaches the Chandrasekhar mass limit. The companion may be a
main-sequence star (WD+MS) or a red-giant star (WD+RG) (Yungelson
et al. \cite{YUN95}; Li et al. \cite{LI97}; Hachisu et al.
\cite{HAC99a}, \cite{HAC99b}; Nomoto et al. \cite{NOM99, NOM03};
Langer et al. \cite{LAN00}; Han \& Podsiadlowski \cite{HAN04};
Chen \& Li \cite{CHENWC07, CHENWC09}; Han \cite{HAN08}; Meng, Chen
\& Han \cite{MENGXC09}; Meng \& Yang \cite{MENGXC09a}; L\"{u}
\cite{LGL09}; Wang et al. \cite{WANGB09a,WANGB09b,WANGB09c}). The
SD model has also been verified by many observations (see Meng \&
Yang \cite{MENGXC09b}). An alternative is the theoretically less
probable double degenerate (DD) model (Iben \& Tutukov
\cite{IBE84}; Webbink \cite{WEB84}), in which a system of two CO
WDs loses orbital angular momentum by means of gravitational wave
radiation and finally merges. The merger remnant may explode if
the total mass of the system exceeds the Chandrasekhar mass limit
(see the reviews by Hillebrandt \& Niemeyer \cite{HN00} and
Leibundgut \cite{LEI00}).

In the single degenerate model, the companion persists after the
supernova explosion. The supernova ejecta collides with the
envelope of and strips some hydrogen-rich material from the
surface of the companion (Marietta et al. \cite{MAR00}; Meng, Chen
\& Han \cite{MENGXC07}; Pakmor et al. \cite{PAK08}). The
stripped-off hydrogen-rich material may manifest itself by means
of narrow H$_{\rm \alpha}$ emission or absorption lines in
later-time spectra of SNe Ia (Chugai \cite{CHU86}; Filippenko
\cite{FIL97}). The amount of the stripped-off material determines
whether or not the narrow hydrogen line can be observed. Marietta
et al. (\cite{MAR00}) ran several high-resolution two-dimensional
numerical simulations of the collision between the ejecta and the
companion. They claimed that about $0.15-0.17$ $M_{\odot}$ of
hydrogen-rich material is stripped from a MS or a subgiant (SG)
companion and 0.5 $M_{\odot}$ from red giant (RG) star. Meng, Chen
\& Han (\cite{MENGXC07}) used a simple analytic method but a more
physical companion model than that used in Marietta et al.
(\cite{MAR00}) to simulate the interaction between SNe Ia ejecta
and companions, and found that the minimum mass of the
stripped-off material from a MS or SG star is $0.035$ $M_{\odot}$.
However, the results of Marietta et al. (\cite{MAR00}) and Meng,
Chen \& Han (\cite{MENGXC07}) did not include confirmation by
observations, i.e., no hydrogen line was detected in nebular
spectra of some SNe Ia and the upper mass limit to the
stripped-off material was set to be 0.02 $M_{\odot}$ (Mattila et
al. \cite{MAT05}; Leonard \cite{LEO07}). Pakmor et al.
(\cite{PAK08}) used a more physical companion model than and a
similar numerical simulation to that of Marietta et al.
(\cite{MAR00}) to recalculate the interaction between the
supernova ejecta and companion. They found results similar to
those of Marietta et al. (\cite{MAR00}). In certain circumstances,
they claimed that these results agree with observations, and hence
that theory does not conflict with observations. However, the
special conditions envisaged by Pakmor et al. (\cite{PAK08})
appear to be unrealistic according to detailed binary population
synthesis results (Meng \& Yang \cite{MENGXC09a}). Based on the
results of Pakmor et al. (\cite{PAK08}), the amount of
stripped-off material is between 0.06 $M_{\odot}$ and 0.16
$M_{\odot}$, which is consistent with the discovery of Marietta et
al. (\cite{MAR00}) and Meng, Chen \& Han (\cite{MENGXC07}) (see
also Meng \& Yang \cite{MENGXC09a}). The results of Pakmor et al.
(\cite{PAK08}) therefore do not resolve the confliction between
theory and observations. Justham et al. (\cite{JUSTHAM09})
proposed that the rotational effect of a CO WD may prevent its
thermonuclear runaway until the accretion phase has ended, which
could produce a RG companion with a low-mass envelope and
reconcile theory and observations. We also suggest that failure to
detect a hydrogen line in nebular spectra of some SNe Ia may imply
that the WD + RG channel is a means of producing SNe Ia. The
amount of hydrogen-rich material obtained by Marietta et al.
(\cite{MAR00}) by means of WD + RG channel is higher than observed
(0.5 $M_{\odot}$), which may be due to the simplistic RG model
used by Marietta et al. (\cite{MAR00}).

In Sect. \ref{sect:2}, we describe our binary evolution model. We
present our evolutionary and binary population synthesis results
in Sect. \ref{sect:3}, and our discussions and conclusions in
Sect. \ref{sect:4}.


\section{Model and physics inputs}\label{sect:2}

Meng \& Yang (\cite{MENGXC09b}) developed a comprehensive
progenitor model for SNe Ia. In the model, the mass-stripping
effect by optically thick wind (Hachisu et al. \cite{HAC96}) and
the effect of a thermally unstable disk were included (Hachisu,
Kato \& Nomoto \cite{HKN08}; Xu \& Li \cite{XL09}). The
prescription of Hachisu et al. (\cite{HAC99a}) for WDs accreting
hydrogen-rich material from their companions was applied to
calculate the WD mass growth. The optically thick wind and the
material stripped-off by the wind were assumed to remove the
specific angular momentum of WD and its companion, respectively.
In Meng \& Yang (\cite{MENGXC09b}), both the WD + MS and WD + RG
scenarios are considered, i.e., Roche lobe overflow (RLOF) begins
at either the MS or RG stage. After the RLOF, WD accretes
hydrogen-rich material from the donor and increases its mass
smoothly. When the mass of the WD reaches 1.378 $M_{\odot}$, the
WD is assumed to explode as a SN Ia. The CO WDs may explode at the
optically thick wind phase or after the optically thick wind while
in either the stable or unstable hydrogen-burning phase or the
dwarf nova phase. They considered more than 1600 different WD
close binary evolution and obtained a dense model grid for SNe Ia.
Based on the comprehensive model, Meng \& Yang (\cite{MENGXC09b})
derived a Galactic birth rate of SNe Ia that is comparable to that
inferred from observations. In the WD + MS channel obtained in
Meng \& Yang (\cite{MENGXC09b}), the companion models are similar
to those in Meng, Chen \& Han (\cite{MENGXC09}). The masses
stripped-off should then be similar to those found in Meng, Chen
\& Han (\cite{MENGXC07}), i.e., higher than $0.035 M_{\odot}$. We
therefore only considered the case of a WD + RG channel, which may
only contribute to a fraction of all observed SNe Ia (see Meng \&
Yang \cite{MENGXC09b} and Wang, Li \& Han \cite{WANGB09c}). All
our methods for calculating the binary evolution and the physics
inputs into the binary evolution calculation are similar to those
in Meng \& Yang (\cite{MENGXC09b}) (see Meng \& Yang
\cite{MENGXC09b} for a detailed description of our method).

   \begin{figure}
   \centering
   \includegraphics[width=60mm,height=80mm,angle=270.0]{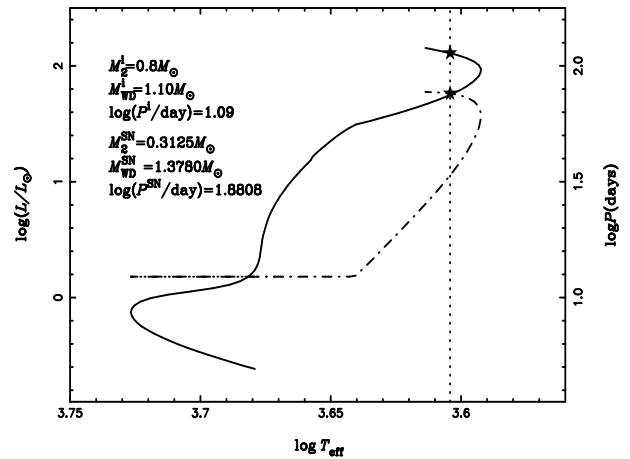}
   \caption{An example of binary evolution calculations. The evolutionary
track of the donor star in HRD is shown as solid curve and the
evolution of orbital period is shown as dot-dashed curve. Dotted
vertical line and asterisks indicate the position where the WD is
assumed to explode as a SN Ia.}
              \label{Fig1}%
    \end{figure}

\section{Results}\label{sect:3}
\subsection{Binary evolution}\label{sect:3.1}
\begin{figure}
   \centering
   \includegraphics[width=60mm,height=80mm,angle=270.0]{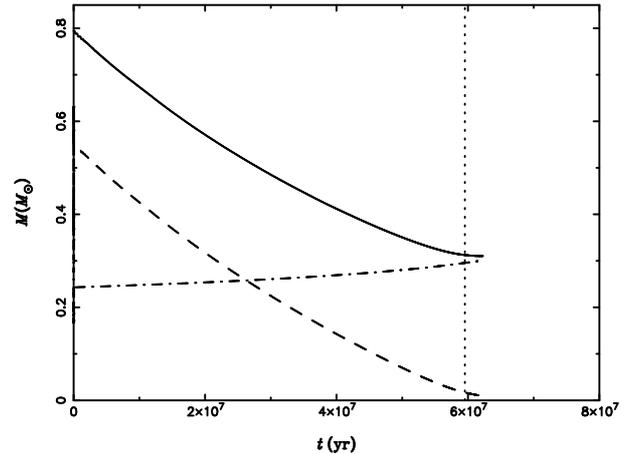}
   \caption{The evolution of secondary mass (sold line), core mass (dot-dashed line)
   and the hydrogen envelope mass (dashed line). Dotted
vertical line indicates the position where the WD is expected to
explode as a SN Ia. The zero point of time is set at the onset of RLOF.} \label{Fig2}%
    \end{figure}

In Fig. \ref{Fig1}, we show the evolutionary track of the donor
star in the Hertzsprung-Russel diagram (HRD, solid line) and the
evolution of the orbital period (dot-dashed line). The initial
parameters of the binary system are also shown in the figure. The
donor star evolves from the zero age main sequence (ZAMS). RLOF
does not begin until the star enters the RG stage. The WD accretes
hydrogen-rich material from the donor and increases its mass to
$M_{\rm WD}^{\rm SN}=1.378 M_{\odot}$, where a SN Ia is assumed to
occur, and $M_{\rm 2}^{\rm SN}=0.3125 M_{\odot}$ and $\log(P^{\rm
SN}/{\rm day})=1.8808$. At this point, the donor star consists of
a helium core of 0.2955 $M_{\odot}$ and a very thin hydrogen
envelope of 0.017 $M_{\odot}$; (we refer to Fig. \ref{Fig2}, which
shows the evolution of donor mass, core mass, and hydrogen
envelope mass, where the definition of the core is identical to
that in Han, Podsiadlowski \& Eggleton (\cite{HAN94}) and Meng,
Chen \& Han (\cite{MENG08})). The envelope is so thin that the
donor star has even left the RG and evolved to become a WD.
Supernova ejecta impacts and strips off hydrogen-rich material
from the envelope. Almost all the envelope material is stripped
off (Marietta et al. \cite{MAR00}). The amount of the mass
stripped-off from the companion should then be lower than 0.017
$M_{\odot}$, which is lower than the upper mass limit of 0.02
$M_{\odot}$ obtained from observations (Mattila et al.
\cite{MAT05}; Leonard \cite{LEO07}). Therefore, no hydrogen line
should be observed in the nebular spectra of this SN Ia, and our
model can be reconciled with observations. This result may imply
that the lack of detection of a hydrogen line in nebular spectra
of some SNe Ia could be evidence that the WD + RG channel
represents a way of producing SNe Ia.

The RG model used by Marietta et al. (\cite{MAR00}) consists of a
helium core of 0.42 $M_{\odot}$ and a thick envelope of 0.56
$M_{\odot}$. The RG model may deviate from reality because the
model was not obtained from a detailed binary evolution
calculation and mass transfer between CO WD and the RG star was
not considered. If mass transfer had been considered, the
structure of the companion would differ significantly from that
used in Marietta et al. (\cite{MAR00}, see Sects. \ref{sect:3.2}
and \ref{sect:4.3}).

\subsection{Binary population synthesis}\label{sect:3.2}
To obtain the distributions of the envelope mass and the core mass
of companion stars from WD + RG channel at the moment of SN
explosion, we performed a detailed Monte Carlo simulation using
Hurley's rapid binary evolution code (Hurley et al. \cite{HUR00,
HUR02}). In the simulation, if a binary system evolves to a WD +
RG stage, and the system is located in the ($\log P^{\rm i},
M_{\rm 2}^{\rm i}$) plane for SNe Ia at the onset of RLOF, we
assume that an SN Ia is then produced. The envelope mass and the
core mass of the WD binary at the moment of SN explosion are
obtained by interpolation in the three-dimensional grid ($M_{\rm
WD}^{\rm i}, M_{\rm 2}^{\rm i}, \log P^{\rm i}$) obtained in Meng
\& Yang (\cite{MENGXC09b}).

In the simulation, we follow the evolution of 40 million sample
binaries. The evolutionary channel was described in Meng \& Yang
(\cite{MENGXC09b}). As for Meng \& Yang (\cite{MENGXC09b}), we
adopted the following input to the simulation: (1) a constant star
formation rate (SFR)over the past 15 Gyr; (2) the initial mass
function (IMF) of Miller \& Scalo (1979); (3) the mass-ratio
distribution is constant; (4) the distribution of separations is
constant in $\log a$ for wide binaries, where $a$ is the orbital
separation; (5) a circular orbit is assumed for all binaries;
(6)the common envelope (CE) ejection efficiency $\alpha_{\rm CE}$,
which denotes the fraction of the released orbital energy used to
eject the CE,  is set to be either 1.0 or 3.0 (see Meng \& Yang
(\cite{MENGXC09b}) for details of the parameter input).

   \begin{figure}
   \centering
   \includegraphics[width=60mm,height=80mm,angle=270.0]{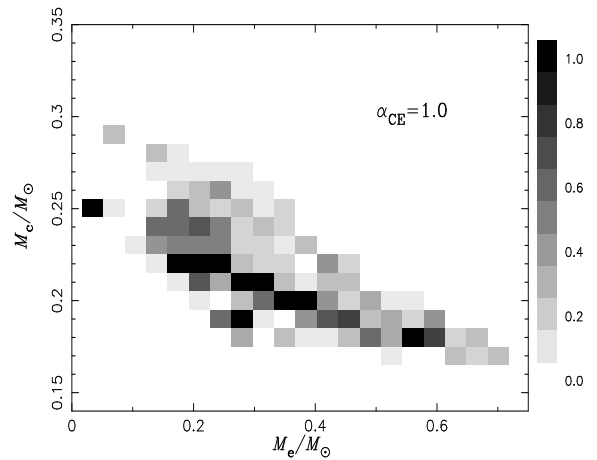}
   \caption{Snapshot distribution of the envelope mass $M_{\rm e}$
   and the core mass $M_{\rm c}$ of companion stars from WD + RG channel at current epoch with $\alpha_{\rm CE}=1.0$.} \label{Fig3}%
    \end{figure}

   \begin{figure}
   \centering
   \includegraphics[width=60mm,height=80mm,angle=270.0]{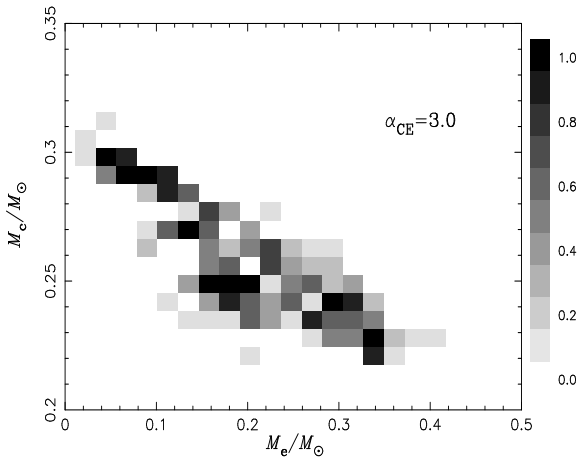}
   \caption{Similar to Fig. \ref{Fig3} but for $\alpha_{\rm CE}=3.0$.} \label{Fig4}%
    \end{figure}

In Figs \ref{Fig3} and \ref{Fig4}, we show current-epoch-snapshot
distributions of the envelope mass and the core mass of companions
produced by the WD + RG channel at the moment of supernova
explosion with different $\alpha_{\rm CE}$\footnote{In the
figures, we do not show the cases of WD + MS because: (i) the
definition of the core for a MS star is incorrect; (ii) the
interaction between supernova ejecta and the companion is complex,
i.e., the stripped-off mass from the companion strongly depends on
both the evolutionary stage of the MS star at the onset of RLOF
and the initial parameters of a WD + MS system (Meng, Chen \& Han
2007); and (iii) the study of WD + MS is beyond the scope of this
paper.}. In these figures, we can discern a clear trend, i.e.,
that core mass decreases in general with envelope mass. This is a
natural consequence of stellar evolution. In addition, varying
$\alpha_{\rm CE}$ may have a significant effect on the
distributions. For $\alpha_{\rm CE}=1.0$, the envelope mass of
companions is mainly between 0.1 $M_{\odot}$ and 0.6 $M_{\odot}$,
while it is lower than 0.4 $M_{\odot}$ for  $\alpha_{\rm CE}=3.0$.
This is because that for a given primordial binary system, high
$\alpha_{\rm CE}$ is indicative of a longer orbital period after
CE ejection. When the RG star fills its Roche lobe, the star then
consists of a low-mass envelope and a high-mass core. As a result,
the lower mass limit of the core mass for $\alpha_{\rm CE}=3.0$ is
higher than that for $\alpha_{\rm CE}=1.0$, while the upper mass
limit of the envelope mass for $\alpha_{\rm CE}=3.0$ is lower than
that for $\alpha_{\rm CE}=1.0$. The core mass is between 0.15
$M_{\odot}$ and 0.3 $M_{\odot}$. After the interaction between
supernova ejecta and RG companion, the RG companion loses almost
its entire envelope (96\% - 98\%) leaving only the core of the
star (Marietta et al. \cite{MAR00}). This may be a channel to
forming single low-mass white dwarf (Justham et al.
\cite{JUSTHAM09}).

However, there are only a few systems with $M_{\rm
e}<0.02M_{\odot}$ in our sample, which means that the binary
sequence considered in this paper may only be able to explain some
individual observations (Mattila et al. \cite{MAT05}; Leonard
\cite{LEO07}).

\section{Discussions and conclusions}\label{sect:4}
\subsection{Age}\label{sect:4.1}
We have found that if a SN Ia originates from the WD + RG channel,
the donor star may almost become a helium WD and only some
hydrogen-rich material remains on top of the helium WD (as little
as 0.017 $M_{\odot}$), which means that the upper limit mass to
the stripped-off from the companion by supernova ejecta is 0.017
$M_{\odot}$. No hydrogen line is then expected in the nebular
spectra of some SNe Ia and our results may explain the conflict
between theory and observation, i.e, theory predicts that the
stripped-off material should be greater than 0.035 $M_{\odot}$
(Marietta et al. \cite{MAR00}; Meng, Chen \& Han \cite{MENGXC07};
Meng \& Yang \cite{MENGXC09b}), while observations indicate that
the upper mass limit of the stripped-off material is 0.02
$M_{\odot}$ (Mattila et al. \cite{MAT05}; Leonard \cite{LEO07}).
No hydrogen line has been detected in nebular spectra of some SNe
Ia may indicate that the progenitors of the observed SNe Ia are
from WD + RG systems. If the observed SNe Ia (SN 2001el, 2005am,
and 2005cl) were produced by the WD + RG channel, they should
originate in an old population. Unfortunately, there is no
constraint on the age of the three SNe Ia. We checked the types of
their host galaxies and found that apart from the host galaxy of
SN 2005cl (MCG-01-39-003), which is a S0 galaxy (Wang
\cite{WANGXF09}; Bufano et al. \cite{BUFANO09}), they are both
spiral galaxies, i.e. the host galaxy of SN 2001el (NGC 1448) is a
Scd galaxy (Wang et al. \cite{WANGLF03}; Wang et al.
\cite{WANGXF06}) and the host galaxy of SN 2001am (NGC 2811) is a
Sa galaxy (Bufano et al. \cite{BUFANO09}). The progenitor of SN
2005cl may therefore belong to an old population, but we are
unable to infer any information about the population of the other
SNe Ia. However, we note that all three SNe Ia are located at the
edge of their host galaxy and that SN 2005cl is even located in a
tail extending from MCG-01-39-003. Is this phenomenon evidence of
an old population? It is possible because halo stars in general
belong to an old population. All three SNe Ia are located at the
edge of their host galaxy may be an observational select effect
because we are more likely to observe a SN Ia at the outskirts of
a host galaxy rather than its inner part. This selection effect
might increase the probability that a SN Ia is produced by the WD
+RG channel with a low-mass envelope being observed.

\subsection{WD + RG system}\label{sect:4.2}
Relative to that of the WD + MS system, the Galactic birth rate of
the WD + RG channel is low (see Meng \& Yang \cite{MENGXC09b} and
Wang, Li \& Han (\cite{WANGB09c})). However, since the Galactic
birth rate of SNe Ia predicted by the model in Meng \& Yang
\cite{MENGXC09b} is lower than that inferred from observations,
the WD + RG channel should be carefully investigated because the
progenitors of some SNe Ia (e.g. SN 2006X and SN 2007on) are
possible WD + RG systems (Patat et al. \cite{PATAT07}; Voss \&
Nelemans \cite{VOSS08}). In addition, some recurrent nova
(belonging to WD + RG) are suggested to be the candidates of SNe
Ia progenitors (Hachisu et al. \cite{HAC99b}; Hachisu \& Kato
\cite{HK06}; Hachisu et al. \cite{HKL07}). In this paper, we even
proposed that the prevalence of the WD + RG channel is why no
hydrogen line was detected in nebular spectra of some SNe Ia,
although the probability of its occurrence is low.

\subsection{Interaction between supernova ejecta and companion}\label{sect:4.3}
Marietta et al. (\cite{MAR00}) performed several high-resolution
two-dimensional numerical simulations of the collision between the
supernova ejecta and companion, and found that a red-giant donor
will lose almost its entire envelope (96\% - 98\%) due to the
impact leaving only the core of the star ($\simeq0.42M_{\odot}$).
The RG star used in Marietta et al. (\cite{MAR00}) consists of a
helium core of $0.42 M_{\odot}$ and an envelope of $0.56
M_{\odot}$, which are not comparable to those obtained by our
simulations (see Figs \ref{Fig3} and \ref{Fig4}). In addition, the
radius of their RG model is 180 $R_{\odot}$, which corresponds to
an orbital period of $\sim900$ days. Too long to compare with our
simulation (see Fig. \ref{Fig1}), the orbital period leads to a
lower envelope binding energy than produced by the model developed
in this paper since the binding energy of the envelope is
determined mainly by the radius of the RG star (Meng, Chen \& Han
\cite{MENG08}). The envelope of the RG model used by Marietta et
al. (\cite{MAR00}) is then more likely to be stripped off and the
amount of material stripped-off by the RG companion in Marietta et
al. (\cite{MAR00}) might be overestimated. A more detailed
numerical simulation of the interaction between supernova ejecta
and an RG companion should therefore be performed by a more
physical companion model than that in Marietta et al.
(\cite{MAR00}).

\subsection{Alternative explanation of the lack of hydrogen}\label{sect:4.4}
The absence of hydrogen lines in the nebular spectra of SNe Ia may
have other explanations. The RG companion with a small
hydrogen-rich envelope may be the results of either a fine-tuning
effect as suggested in this paper, or a physical process. For
example, Justham et al. (\cite{JUSTHAM09}) suggested that a
rotational effect of WD could prevent the thermonuclear runaway
occurring until the accretion phase has ended, which could also
produce a RG companion with a low-mass envelope. Rotation may also
increase the probability of a SN Ia from WD +RG channel with
low-mass envelope being observed. An alternative explanation of
the lack of hydrogen is that the amount of stripped-off material
might has been dramatically overestimated as discussed above.
\\

Based on the discussions above, further attempts to observe
hydrogen lines in nebular spectra of SNe Ia are encouraged to
check our suggestion. The WD + RG system may also be an origin of
single low-mass white dwarfs.

\begin{acknowledgements}
We are grateful to the anonymous referee for his/her constructive
suggestions improving this manuscript greatly. This work was
supported by Natural Science Foundation of China under grant no.
10963001.
\end{acknowledgements}


\begin{thebibliography}{}
\bibitem[2004]{BRA04}
Branch D., 2004, Nature, 431, 1044
\bibitem[2009]{BUFANO09}
Bufano F., Immler S., Turatto M. et al., 2009, ApJ, 700, 1456
\bibitem[2007]{CHENWC07}
Chen W., Li X., 2007, ApJ, 658, L51
\bibitem[2009]{CHENWC09}
Chen W., Li X., 2009, ApJ, 702, 686
\bibitem[1986]{CHU86}
Chugai N.N., 1986, SvA, 30, 563
\bibitem[1997]{FIL97}
Filippenko A.V., 1997, ARA\&A, 35, 309
\bibitem[1996]{HAC96}
Hachisu I., Kato M., Nomoto K., 1996, ApJ, 470, L97
\bibitem[1999a]{HAC99a}
Hachisu I., Kato M., Nomoto K. et al., 1999a, ApJ, 519, 314
\bibitem[1999b]{HAC99b}
Hachisu I., Kato M., Nomoto K., 1999b, ApJ, 522, 487
\bibitem[2006]{HK06}
Hachisu I., Kato M., 2006b, ApJ, 651, L141
\bibitem[2007]{HKL07}
Hachisu I., Kato M., Luna G.J.M., 2007, ApJ, 659, L153
\bibitem[2008]{HKN08}
Hachisu, I., Kato, M., Nomoto, K., 2008, ApJ, 679, 1390
\bibitem[1994]{HAN94}
Han Z., Podsiadlowski P., Eggleton P.P., 1994, MNRAS, 270, 121
\bibitem[2004]{HAN04}
Han Z., Podsiadlowski Ph., 2004, MNRAS, 350, 1301
\bibitem[2008]{HAN08}
Han Z., 2008, ApJ, 677, L109
\bibitem[2000]{HN00}
Hillebrandt W., Niemeyer J.C., 2000, ARA\&A, 38, 191
\bibitem[2000]{HUR00}
Hurley J.R., Pols O.R., Tout C.A., 2000, MNRAS, 315, 543
\bibitem[2002]{HUR02}
Hurley J.R., Tout C.A., Pols O.R., 2002, MNRAS, 329, 897
\bibitem[1984]{IBE84}
Iben I., Jr., Tutukov A.V., 1984, ApJS, 54, 335
\bibitem[2009]{JUSTHAM09}
Justham S., Wolf C., Podsiadlowski P., Han, Z., 2009, A\&A, 493,
1081
\bibitem[2000]{LAN00}
Langer N., Deutschmann A., Wellstein S. et al., 2000, A\&A, 362,
1046
\bibitem[2000]{LEI00}
Leibundgut B., 2000, A\&ARv, 10, 179
\bibitem[2007]{LEO07}
Leonard D.C., 2007, ApJ, 670, 1275
\bibitem[1997]{LI97}
Li X.D., van den Heuvel E.P.J., 1997, A\&A, 322, L9
\bibitem[2009]{LGL09}
L\"{u}, G., Zhu, C. Wang, Z., Wang, N., 2009, MNRAS, 396, 1086
\bibitem[2000]{MAR00}
Marietta E., Burrows A., Fryxell B., 2000, ApJS, 128, 615
\bibitem[2005]{MAT05}
Mattila S., Lundqvist P., Sollerman J. et al., 2005, A\&A, 443,
649
\bibitem[2007]{MENGXC07}
Meng X., Chen X., Han Z., 2007, PASJ, 59, 835
\bibitem[2008]{MENG08}
Meng X., Chen X., Han Z., 2008, A\&A, 487, 625
\bibitem[2009]{MENGXC09}
Meng X., Chen X., Han Z., 2009, MNRAS, 395, 2103
\bibitem[2010a]{MENGXC09a}
Meng X., \& Yang W., 2010a, MNRAS, 401, 1118
\bibitem[2010b]{MENGXC09b}
Meng X., \& Yang W., 2010b, ApJ, 710, 1310
\bibitem[1984]{NTY84}
Nomoto K., Thielemann F-K., Yokoi K., 1984, ApJ, 286, 644
\bibitem[1999]{NOM99}
Nomoto K., Umeda H., Hachisu I. et al., 1999, in Truran J.,
Niemeyer T., eds, Type Ia Suppernova : Theory and
Cosmology.Cambridge Univ. Press, New York, p.63
\bibitem[2003]{NOM03}
Nomoto K., Uenishi T., Kobayashi C. et al., 2003, in Hillebrandt
W., Leibundgut B., eds, From Twilight to Highlight: The Physics of
supernova, ESO/Springer serious ``ESO Astrophysics Symposia''
Berlin: Springer, p.115
\bibitem[2008]{PAK08}
Pakmor R.; R\"{o}opke F.K.; Weiss A.; Hillebrandt W., 2008, A\&A,
489, 943
\bibitem[2007a]{PATAT07}
Patat F. et al., 2007, Science, 317, 924
\bibitem[1999]{PER99}
Perlmutter S., Aldering G., Goldhaber G. et al., 1999, ApJ, 517,
565
\bibitem[1998]{REI98}
Riess A.G., Filippenko A.V., Challis P. et al., 1998, AJ, 116,
1009
\bibitem[2008]{VOSS08}
Voss R., \& Nelemans G., 2008, Nature, 451, 802
\bibitem[1995]{YUN95}
Yungelson L., Livio M., Tutukov A. et al., 1995, ApJ, 447, 656
\bibitem[2003]{WANGLF03}
Wang L., Baade D., H\"{o}flich P. et al., 2003, ApJ, 591, 1110
\bibitem[2006]{WANGXF06}
Wang X., Wang L., Pain R. et al. 2006, ApJ, 645, 488
\bibitem[2009]{WANGXF09}
Wang X., Li W., Filippenko A.V. et al.  2009, ApJ, 697, 380
\bibitem[2009a]{WANGB09a}
Wang, B., Meng, X., Chen, X., Han, Z., 2009a, MNRAS, 395, 847
\bibitem[2009b]{WANGB09b}
Wang, B., Chen, X., Meng, X., Han, Z., 2009b, ApJ, 701, 1540
\bibitem[2010]{WANGB09c}
Wang, B., Li, X., Han, Z., 2010, MNRAS, 401, 2729
\bibitem[1984]{WEB84}
Webbink R.F., 1984, ApJ, 277, 355
\bibitem[1973]{WI73}
Whelan J. \& Iben I., 1973, ApJ, 186, 1007
\bibitem[2009]{XL09}
Xu X. \& Li X., 2009, A\&A, 495, 243

\end{thebibliography}
\end{document}